\def\year{2022}\relax
\documentclass[letterpaper]{article} % DO NOT CHANGE THIS
\usepackage{aaai21x}  % DO NOT CHANGE THIS
\usepackage{times}  % DO NOT CHANGE THIS
\usepackage{helvet} % DO NOT CHANGE THIS
\usepackage{courier}  % DO NOT CHANGE THIS
\usepackage[hyphens]{url}  % DO NOT CHANGE THIS
\usepackage{graphicx} % DO NOT CHANGE THIS
\usepackage{natbib}  % DO NOT CHANGE THIS AND DO NOT ADD ANY OPTIONS TO IT
\usepackage{caption} % DO NOT CHANGE THIS AND DO NOT ADD ANY OPTIONS TO IT
\usepackage{booktabs}

\usepackage{amsmath}
\usepackage{amssymb}
\usepackage{makecell}
\usepackage{float}
\usepackage{subcaption}
\usepackage{hyperref}
\usepackage{hyphenat}
\usepackage{mathptmx}

\usepackage[english]{babel}
\usepackage{csquotes}
\usepackage{xcolor}
\usepackage{verbatim}
\usepackage{booktabs}
\usepackage{breqn}

 \usepackage{booktabs}
\usepackage{caption}
\title{The Amplification Paradox in Recommender Systems}
\author{
Manoel Horta Ribeiro, Veniamin Veselovsky, Robert West\\
}
\affiliations{EPFL\\
manoel.hortaribeiro@epfl.ch, veniamin.veselovsky@epfl.ch, robert.west@epfl.ch}
\begin{document}

\maketitle

\begin{abstract}
Automated audits of recommender systems found that blindly following recommendations leads users to increasingly partisan, conspiratorial, or false content. 
At the same time, studies using real user traces suggest that  recommender systems are not the primary driver of attention toward extreme content; on the contrary, such content is mostly reached through other means, e.g., other websites.
In this paper, we explain the following apparent paradox: \emph{if the recommendation algorithm favors extreme content, why is it not driving its consumption?}
With a simple agent-based model where users attribute different utilities to items in the recommender system, we show through simulations that the collaborative-filtering nature of recommender systems and the nicheness of extreme content can resolve the apparent paradox: 
although blindly following recommendations would indeed lead users to niche content,
users rarely consume niche content when given the option because it is of low utility to them, which can lead the recommender system to deamplify such content.
Our results call for a nuanced interpretation of ``algorithmic amplification'' and highlight the importance of modeling the utility of content to users when auditing recommender systems.
Code available: \url{https://github.com/epfl-dlab/amplification_paradox}.
\end{abstract}

\section{Introduction}\label{sec:intro}
On social media platforms, recommender systems bridge the gap between content creators and regular users. 
On the one hand, they enable users to navigate through vast content catalogs effortlessly.
On the other hand, they help content creators find an audience. 
As recommender systems become pervasive, scholars~\cite{whittaker2021recommender}, the media~\cite{nyt1}, and even the general public~\cite{regrets} have criticized the misalignment between what recommender systems optimize for and the goals of users and society.
For instance, on YouTube, one of the world’s largest social media platforms, the recommender system is perceived to amplify inappropriate or fringe content (e.g., conspiracy theories). 
Motivated by the concern of algorithmic amplification, recent studies using sock puppets have audited YouTube's recommender system, showing that watching videos related to misinformation or pseudoscience causes YouTube to recommend more such content~\cite{hussein2020measuring,papadamou2022just,haroon2022youtube,brown2022echo}.

However, recent work using real navigation logs complicates this narrative, showing that YouTube’s recommender system is not the primary driver of attention toward extreme content~\cite{hosseinmardi2021examining, chen2022subscriptions}.
On the contrary, extreme content is often reached through other websites and is not frequently present in long algorithmically driven watching sessions. 
These findings are aligned with the ``supply-and-demand'' hypothesis for the rise of fringe content on platforms like YouTube~\cite{munger2022right}: ``problematic'' content thrives because people want to consume it, and social media affordances (e.g., the ease of distributing videos to niche audiences and monetizing it) allow this demand to be met.

Here, we propose an agent-based model that explains the central paradox emerging from the aforementioned literature, which we name the ``amplification paradox:'' \emph{if the recommendation algorithm favors extreme content, why is it not driving its consumption?}
While our model is simpler than the recommender systems in production on platforms like YouTube, it shows how the collaborative\hyp filtering nature of recommender systems and the nicheness of extreme content can, by themselves, explain the contradicting observations in previous work (i.e., \emph{the algorithm favors extreme content} vs.\ \emph{the algorithm does not drive the consumption of extreme content}). 
The reason is that,
although blindly following recommendations would indeed lead users to niche content,
users rarely consume niche content when given the option because it is of low utility to them, which can lead the recommender system to deamplify such content.

These results have key implications.
First, they suggest that algorithmic audits on recommender systems are of limited utility in determining the prevalence of phenomena like radicalization, rabbit holes, and filter bubbles \emph{if they do not model how users interact with algorithms.}
To meaningfully represent reality, algorithmic audits ought to model user preferences, as users do not blindly follow recommendations~\cite{lee2022algorithmic}.
Second, they indicate the dynamics of extreme or harmful content (e.g., QAnon conspiracy) within algorithmically driven platforms may be explained, at least in part, by the nicheness of the content, as our model considers nothing but the popularity and co-consumption patterns of different items.
Third, they highlight the need for nuance around the notion of ``algorithmic amplification,'' which we argue should consider the utility of content towards users.

\section{Agent-based model}

Our model captures three key ingredients present in online platforms like YouTube:
1) the recommender systems suggest items that similar users have consumed;
2) different topics appeal to different audiences; and 
3) users consume content according to their internal preferences.

\begin{figure}[t]
    \centering
    \includegraphics[width=\linewidth]{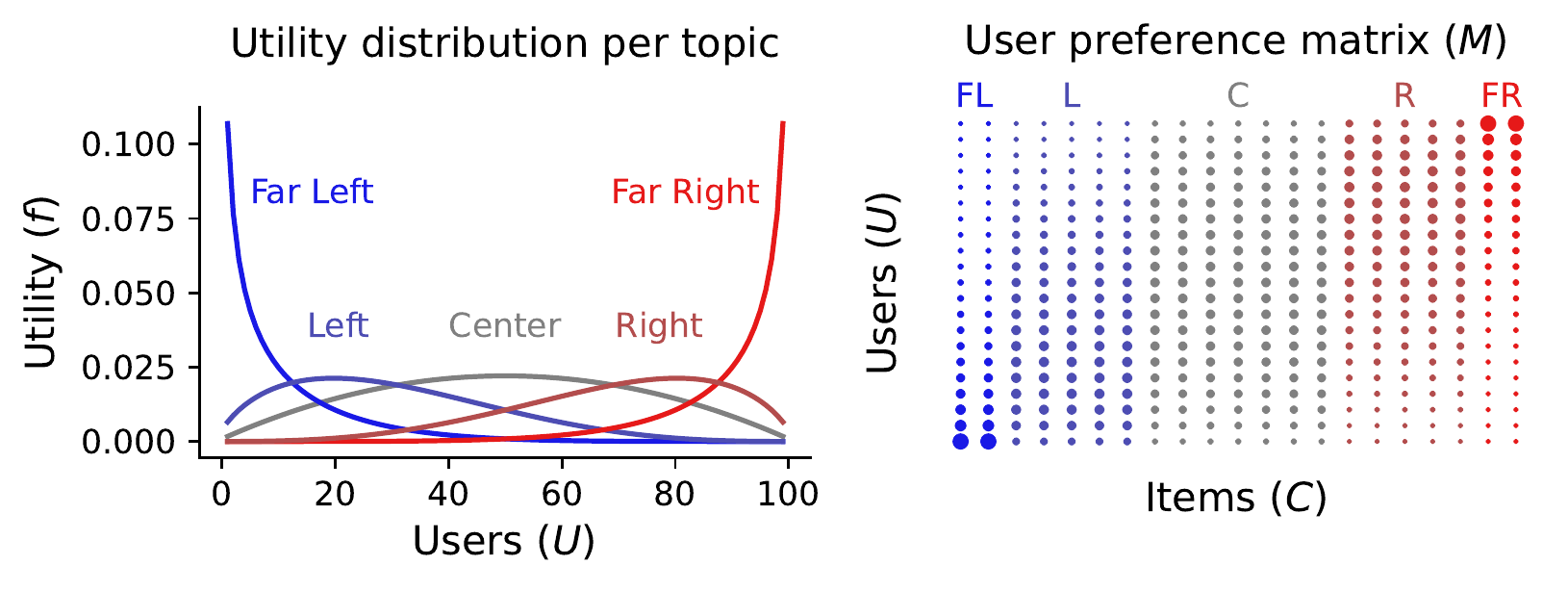}
    \caption{On the left, we depict the scenario considered: five topics, each corresponding to a political position that appeals differently to users.
    On the right, we  depict the user preference matrix $M$, where the size of the dots represents the utility of a user--item pair. Users are ordered from the highest index (top; most right-leaning) to the lowest (bottom; most left-leaning), and items are ordered according to their topic.
    }
    \label{fig:example}
\end{figure}

\noindent
\textbf{User preferences.}
We consider the scenario commonly used in the literature [e.g., ~\citet{haroon2022youtube}, \citet{hosseinmardi2021examining}], with five topics: ~\emph{Far Left}~(\textbf{FL}), \emph{Left} (\textbf{C}), \emph{Center} (\textbf{C}), \emph{Right} (\textbf{R}), \emph{Far Right} (\textbf{FR}) that appeal differently to individuals across the political spectrum.
We illustrate our \emph{desiderata} in Fig.~\ref{fig:example}~(left).
Considering users that range from the most left-leaning (user 1 in the figure) to very right-leaning (user 100), items from a topic are high-utility to users whose political views are well-aligned, e.g., items from the \emph{Far Left} topic are high-utility to low-index users and low-utility to high-index users. 
Further, the more extreme a topic, the more its utility distribution is concentrated among a few users.
Last, items belonging to the same topic are indistinguishable, i.e., each item of a given topic (e.g., \emph{Left}) has the same utility for a given user.

We operationalize this scenario by constructing a matrix $M$ of dimensions  $|U| \times |C|$ capturing user preferences, i.e., each element $m_{ij}$ captures the utility of item $j$ to user $i$. To flexibly model $m_{ij}$, we use the (scaled) probability mass function of the beta-binomial distribution:
\begin{equation}\label{eq:m}
\small
m_{ij} = \gamma_j \, 
\overbrace{
{|U| \choose i} \, \frac{\mathrm{B}(i + \alpha_j, |U| - i + \beta_j)}{\mathrm{B}(\alpha_j, \beta_j)}
}^{\text{Beta-binomial PMF}},
\end{equation}
where $\mathrm{B}$ is the beta function, $\alpha_j$ and  $\beta_j$ are \emph{concentration parameters} that control the shape of the curve, and $\gamma_j$ is a \emph{scale parameter} that determines the area under the curve (when $\gamma_j$ equals 1, so does the area under the curve). 
For each topic, we consider items that all share the same parameters, e.g., the topic \textit{Left} has $\eta_L$ items each associated with parameters $\alpha_{L}$,  $\beta_{L}$, and $\gamma_{L}$.
We illustrate an $M$ matrix constructed as described above in Fig.~\ref{fig:example}~(right), where each dot represents the utility $m_{ij}$ associated with a user--item pair.

\vspace{1.3mm}
\noindent
\textbf{Recommender system.}
Let a collaborative\hyp filtering recommender system (CFRS) serve items from an item catalog $C$ to users $U$. 
The CFRS shows items to users and records which items users consume.
We represent the input of the CFRS as a $|U| \times |C|$ matrix $S$, where each element $s_{ij}$ equals $1$ if user $i$ has consumed item $j$ in the past, and $0$ otherwise.
Let   $N^w_U(i)$ be the set of the $w$ users most similar to user $i$ in matrix $S$ according to the cosine similarity ($\text{cos}$). 
We estimate a score $\hat{s}_{ij}$ for the user--item pair $\langle i, j\rangle $ as

\begin{equation}
\small
    \hat{s}_{ij} =  \frac{
\sum_{k \in N_U^w(i)} \text{cos}(s_{i*}, s_{k*}) \, s_{kj}}
{\sum_{k \in N_U^w(i)} \text{cos}(s_{i*}, s_{k*})},
\end{equation}
where $s_{i*}$ is the $i$-th row of $S$, containing user $i$'s records.

\vspace{1.3mm}
\noindent
\textbf{Interaction.}
Let $z_i \sim \text{Poisson}(\lambda)$ be the number of interaction rounds of user $i$ with the recommender system.
At each round, the recommender system provides the user with a set of $v$ items that the user has not yet consumed. 
We consider two ways in which users select an item among recommendations given to them. Either they choose items uniformly at random, i.e., disregarding the utility of the items (``random selection''), or they select an item $j$ at random with a probability proportional to the item's utility $m_{ij}$ (``utility-informed selection'').

\vspace{1.3mm}
\noindent
\textbf{Relative utility.} 
Let the relative utility $r_{iq}$ be the percentage of content belonging to topic $q$ that user $i$ would consume if they choose items from the whole catalog at random with probability proportional to each item's utility, i.e.,
\begin{equation}
\small
    r_{iq} = \sum_{j \in q} m_{ij} \Big{/} 
    \sum_{j \in  C} m_{ij}.
\end{equation}
This is similar to  \citeauthor{chang2022recommend}'s \citeyear{chang2022recommend} ``organic model,'' a counterfactual that simulates consumption without a recommender system. 
A topic is ``amplified'' [``deamplified''] by the recommender system for a user if user consumption of the topic is above [below] its relative utility; e.g., if the relative utility of the topic \textit{Left} for user $i$ equals 25\%, but 50\% of the items $i$ consumed were from the topic, the topic is said to be amplified by the recommender system.

\begin{figure*}[t]
    \centering

    \includegraphics[scale=0.5]{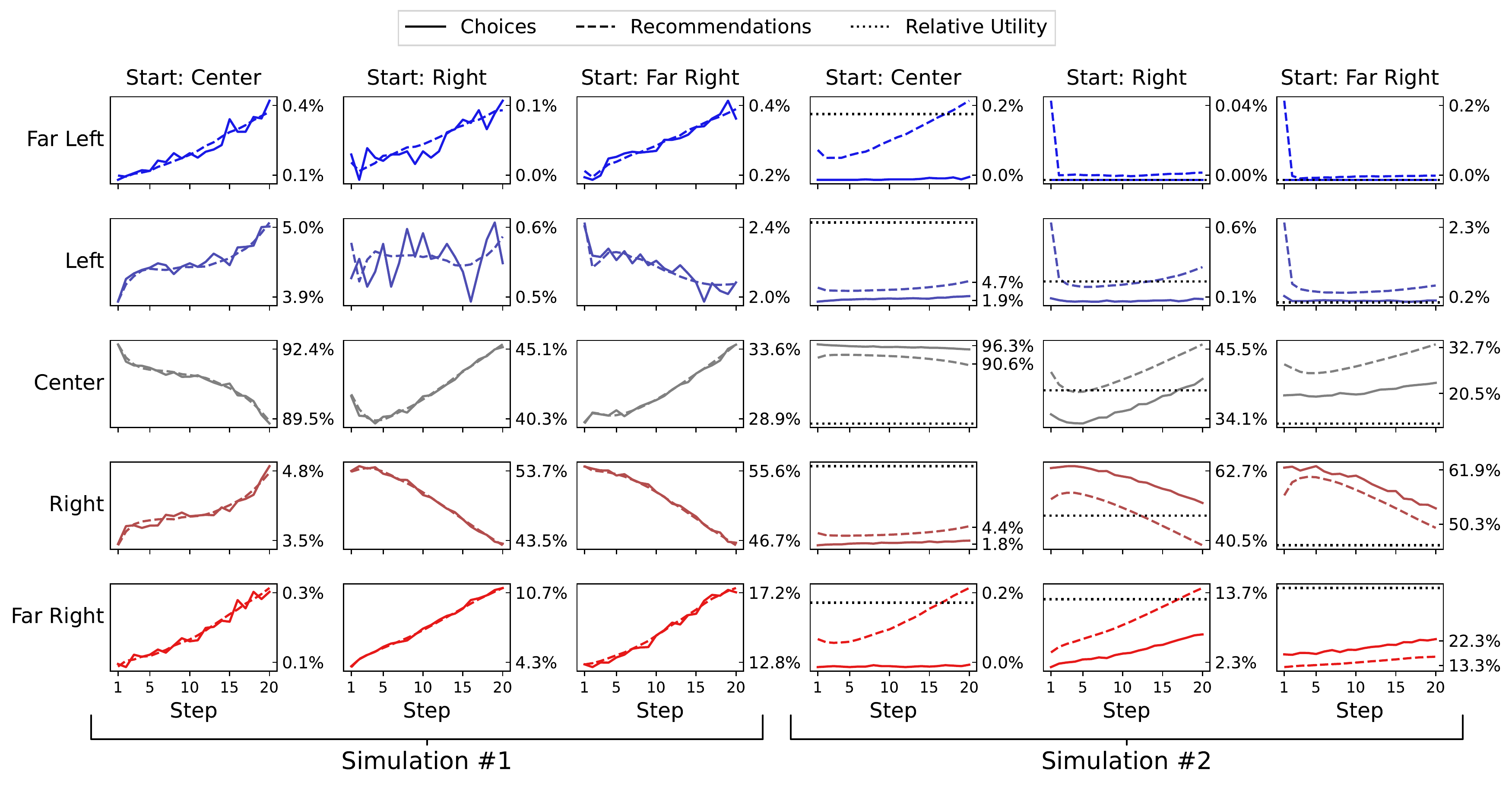}

    \caption{Simulation results. The $y$-axes in the plots show the percentage of times users chose (solid line) or were recommended (dashed line) an item of a specific topic (one per row) in different starting conditions (i.e., what video we initialize the user history with; one per column). The $x$-axis depicts the number of steps in the simulation. For the second simulation, we also show each topic's relative utility (dotted line), a counterfactual estimate simulating consumption without a recommended system, i.e., if users choose from the whole catalog of items with probability proportional to each item's utility to them. We omit starting conditions \textit{Left}/\textit{Far Left} as they are symmetrical to \textit{Right}/\textit{Far Right}, $y$ scales differ per subplot. }
    \label{fig:exp}
\end{figure*}

\vspace{1.3mm}
\noindent
\textbf{Simulation procedure.}
First, in the \emph{burn-in phase}, we populate the matrix $S$.
Until all users have carried out all interactions with the recommender, we
1) sample a random user $i$ who has interacted with the recommender system fewer than $z_{i}$ times, and 2) let $i$ interact with the recommender system with utility-informed selection.
Second, in the \emph{measurement phase}, we quantify how new users would receive recommendations given an already\hyp populated matrix $S$. We
1)~sample a random user $i$ and temporarily erase their corresponding row in the CFRS matrix, i.e., we set $s_{i*} = \mathbf{0}$,
2)~add one item to the user vector, creating a starting condition that varies depending on the simulation,
3)~let $i$ interact with the recommender system (the selection procedure is either random or utility-informed, depending on the simulation).

\vspace{1.3mm}
\noindent
\textbf{Parameter summary.}
Altogether, our model has the following parameters: the number of users $|U|$ and items $|C|$, the number of recommendations $v$ given and nearest neighbors $w$ used by the recommender system, the parameter $\lambda$ governing the number of times each user interacts with the recommender system, and the number of topics $|T|$ and, for each topic $q$,  parameters $\alpha_q$, $\beta_q$, $\gamma_q$,  and $\eta_q$.

\section{Simulations}

\begin{table}[t]
\centering
\footnotesize
    \caption{
    Parameters used in our simulation, note that for $\alpha$, $\gamma$, and $\eta$, we list the parameter associated with each topic (\textbf{L}, \textbf{CL}, \textbf{C}, \textbf{CR}, \textbf{R}). We omit $\beta$ as the parameters used are symmetrical, e.g., $\alpha_L = \beta_R$, $\alpha_{CL} = \beta_{CR}$.
}\label{tab:params}
\begin{tabular}{ll|ll|ll}
\toprule
$|T|$: & 5   & $\lambda$: & 60 & $\alpha$: & 1, 1, 1.3, 5, 16      \\
$|U|$: & 600 & $v$:      & 20 & $\gamma$: & 1, 1.2, 1.5, 1.2, 1   \\
$|C|$: & 600 & $w$:      & 10 & $\eta$:   & 75, 125, 200, 125, 75 \\ \bottomrule
\end{tabular}

\end{table}

% \textbf{Model parameters.} 
% We can run simulations with this agent-based model after setting the number of users $|U|$ and items $|C|$, the number of recommendations $r$ given and nearest neighbors $m$ used by the recommender system, the parameter $\lambda$ governing the number of times each user interacts with the recommender system, and the number of topics $|T|$ and, for each topic $q$,  parameters $\alpha_q$, $\beta_q$, $\gamma_q$,  and $\eta_q$.

We conduct two simulations that attempt to explain the amplification paradox.
Both share the same parameters (see Tab.~\ref{tab:params}; discussed in Sec.~\ref{sec:disc}) and the same burn-in phase.

\vspace{1.3mm}
\noindent
\textbf{Simulation \#1} examines what is recommended after users consume items from a topic and then blindly follow recommendations, similar to how recent studies audit recommender systems [e.g., \citet{brown2022echo}].
If we observe that the algorithm favors niche content, we will have reached similar conclusions to previous work [assuming ``extreme'' content is niche, which previous work supports, see~\citet{ribeiro2020auditing}].
In our agent-based model, we operationalize this simulation by, in the measurement phase, adding a random item to the user vector in step \#2 and interacting with the recommended through random selections in step \#3.
We then analyze the percentage of times items from each topic are recommended/chosen depending on the topic of the item topic added to the user history.

\vspace{1.3mm}
\noindent
\textbf{Simulation \#2} examines how topics are recommended and consumed when users follow their preferences. 
If the algorithm does not drive the consumption of extreme topics, as indicated by studies analyzing real user traces~\cite{hosseinmardi2021examining,chen2022subscriptions}, we would expect that these topics are not systematically amplified.  
In our agent-based model, we operationalize this simulation by, in the measurement step,
adding an item of the topic of the highest utility to the randomly selected user in step \#2 and interacting with the recommended through utility-informed selections in step \#3. 
Again, we estimate the percentage of times each topic is recommended/chosen.

\subsection{Results}

We present the results of simulations \#1 and \#2 in Fig.~\ref{fig:exp} (left and right, respectively). The figure reads like a table.
Each row shows the  percentage of times a topic was  recommended and chosen by users that were initialized with items from different topics, each in a column. We
show only three initial conditions (\emph{Center}, \emph{Right}, and \emph{Far Right}) as topics are symmetrical: e.g., the occurrence of \emph{Right} items under the starting condition \emph{Far Left} equals the occurrence of \emph{Left} items under the starting condition \emph{Far Right}, etc. 

In simulation \#1, we find that no matter where users start, they become increasingly exposed to content in the \emph{Far Right} and the \emph{Far Left}, the most niche and ``extreme'' of topics, e.g., users that start with one \emph{Far Right} video in their history (third column) go from having roughly 13\% of recommended videos belonging to the \emph{Far Right} topic when interacting with the recommender system for the first time in step 1 to having around 17\% in step 20.
This is similar to what recent studies found when auditing the YouTube recommender system~\cite{haroon2022youtube,brown2022echo} with bots. Note that as the selection here is random, the fraction of topics recommended and chosen are very similar.

In simulation \#2,  we also depict the \emph{relative utility} of each topic to users in each initial condition as a horizontal dotted line in each plot.
We find that the \emph{Far Left} and \emph{Far Right} topics (in the first and the fifth row, respectively) are rarely recommended to, and chosen by, users who start in the \emph{Center} initial condition (fourth column). 
Considering users that start on the \emph{Far Right} initial condition (sixth column), we see that \emph{Far Right} content is not recommended or chosen substantially more than in Simulation \#1, and \emph{Left} items are seldom recommended and never chosen. 
Most important, across all starting conditions, extreme content is never chosen above the relative utility of the items to users in the starting condition. As the users are randomly sampled in the experiment, we more generally state that, on average, \textit{Far Right} and \textit{Far Left} items are deamplified by the recommender system.
This is in accordance with the analyses of real user traces from previous work, e.g., \citet{hosseinmardi2021examining} have found that consumers of extreme content do not consume more extreme content deep into long algorithmically driven viewing sessions.

\section{Discussion}
\label{sec:disc}

Our first simulation shows that the most extreme topics (\textit{Far Left} and \textit{Far Right}) are increasingly recommended when blindly following recommendations, similar to what \citet{haroon2022youtube} and other recent audits observe.
However, when users choose items based on their preferences, as in our second simulation, we find that extreme topics are \textit{deamplified} by the recommender system, i.e., users consume these topics less than they would have in the absence of a recommender system. 
This is aligned with empirical studies with real navigation logs~\cite{hosseinmardi2021examining, chen2022subscriptions} that have found that the recommender system is not a key driver of extreme content.
Since users do not meet their demand for this content through recommendations, it is only natural that they resort to subscriptions or other websites to find it.
Thus, we provide a simple potential explanation for the amplification paradox: 
although blindly following recommendations would indeed lead users to niche content, users rarely consume niche content when given the option because it is of low utility to them, which can lead the recommender system to deamplify such content.
Importantly, our findings have nothing to do with how ``ideologically extreme'' a topic is \textit{but how niche it is}. Thus, we might observe this same behavior with harmless niche content (e.g., Japanese carpentry), which may appeal to a specific group.

\citet{metaxa2021auditing} define an algorithm audit as \textit{``a method
of repeatedly and systematically querying an algorithm with inputs and observing the corresponding outputs in order to draw inferences about its opaque inner workings.''} 
This methodology is appropriate to audit ``single-round'' interactions between humans and algorithms, e.g., when \citet{buolamwini2018gender} show how commercial gender classification algorithms  systematically misclassify darker-skinned women.
However, recent audits of the YouTube recommender system try to uncover phenomena that, like ``echo chambers,''  arise from multiple interactions between humans and algorithms \textit{without realistically modeling the human side of the interaction}~\cite{haroon2022youtube,brown2022echo}.
Our agent-based model illustrates how factoring in user preferences can yield substantially different results, and, therefore, it follows that audits on YouTube are of limited utility in determining the prevalence of phenomena like radicalization, echo chambers, etc., insofar as they do not realistically model how users interact with recommender systems [a growing research topic, e.g., see \citet{lee2022algorithmic} and \citet{shin2020users}].

The limitations of algorithmic audits on YouTube reflect a broader issue with the notion of ``algorithmic amplification.'' 
While the term is increasingly present in the regulatory debate [see \citet{whittaker2021recommender}], experts have pointed out that it is ambiguous~\cite{amp1} and that enforcing laws around it is challenging~\cite{keller2021amplification}.
In our model, we adopt a ``utility-based'' notion of algorithmic amplification; we consider that a topic is amplified if it is systematically consumed by users attributing low utility to it.
This perspective, currently not present in the regulatory debate~\cite{keller2021amplification}, can help stakeholders more clearly understand recommender systems.

% parameters simplicity purpose
A possible criticism of the work at hand is that this is an exceedingly simple model and that evaluations were not thorough (e.g., we did not examine the model with various parameters). 
We argue that these flaws do not undermine our results, as the purpose of the model is to provide a possible  explanation for seemingly contradictory results in the existing literature and not to create a realistic model of how users interact with the YouTube recommender system. In a sense, this paper is analogous to an ``existence proof,'' showing that \textit{there exists} a simple model that, parametrized a certain way (which we argue is reasonable), can explain the results in the literature.
Nonetheless, extending the present model to be more realistic may be a worthy pursuit. Similar to how we can reason about possible answers to the ``amplification paradox'' given our simple model, other models that take into account how user preferences are shaped by the recommender system~\cite{ben2018game,cotter2019playing} or how the recommender system creates incentives to produce specific kinds of content~\cite{kalimeris2021preference} may help guide empirical work trying to understand the impact of recommender systems on society.

\vspace{1.3mm}
\noindent
\textbf{Ethical considerations.}
We do not foresee a negative societal impact coming from this research, which, on the contrary, may help improve algorithmic audits of recommender systems like YouTube, TikTok, and Instagram.

% Bibliography
{
\bibliography{00MainPaper}
}
\appendix

\end{document}